\documentclass[10pt,twocolumn]{article}

\usepackage[utf8]{inputenc}
\usepackage[T1]{fontenc}
\usepackage{amsmath, amssymb}
\usepackage{graphicx}
\usepackage{tikz}
\usetikzlibrary{shapes, arrows, positioning, calc, fit, shadows, backgrounds}
\usepackage{cite}
\usepackage{url}
\usepackage{hyperref}
\usepackage{mathptmx}        
\usepackage{booktabs}
\usepackage{caption}
\usepackage{subcaption}
\usepackage{array}

\usepackage{geometry}
\title{Revenue-Sharing as Infrastructure: A Distributed Business Model for Generative AI Platforms}
\author{
  \textbf{Ghislain Dorian Tchuente Mondjo}\\
  \textit{University of Yaoundé I, Yaoundé, Cameroon}\\
  \texttt{tchuente.mondjo@gmail.com}
}
\date{}
\begin{document}

\twocolumn[
  \begin{@twocolumnfalse}
    \maketitle
    \begin{abstract}
      Generative AI platforms (Google AI Studio, OpenAI, Anthropic) provide infrastructures (APIs, models) that are transforming the application development ecosystem. Recent literature distinguishes three generations of business models: a first generation modeled on cloud computing (pay-per-use), a second characterized by diversification (freemium, subscriptions), and a third, emerging generation exploring multi-layer market architectures with revenue-sharing mechanisms. Despite these advances, current models impose a financial barrier to entry for developers, limiting innovation and excluding actors from emerging economies. This paper proposes and analyzes an original model, "Revenue-Sharing as Infrastructure" (RSI), where the platform offers its AI infrastructure for free and takes a percentage of the revenues generated by developers' applications. This model reverses the traditional upstream payment logic and mobilizes concepts of value co-creation, incentive mechanisms, and multi-layer market architecture to build an original theoretical framework. A detailed comparative analysis shows that the RSI model lowers entry barriers for developers, aligns stakeholder interests, and could stimulate innovation in the ecosystem. Beyond its economic relevance, RSI has a major societal dimension: by enabling developers without initial capital to participate in the digital economy, it could unlock the "latent jobs dividend" in low-income countries, where mobile penetration reaches 84\%, and help address local challenges in health, agriculture, and services. Finally, we discuss the conditions of feasibility and strategic implications for platforms and developers.
    \end{abstract}
    \vspace{0.3cm}
    \vspace{0.5cm}
  \end{@twocolumnfalse}
]

\section{Introduction}

The advent of generative AI (GenAI) has profoundly transformed the digital platform landscape. Models such as GPT-4 (OpenAI), Gemini (Google), or Claude (Anthropic) have given rise to a new ecosystem where providers of foundation models offer their infrastructure to third-party developers. These developers in turn create applications for end users, forming a multi-level value chain.

In this context, the question of the optimal business model for these platforms remains largely open. Model providers must indeed arbitrate between several objectives: maximizing their revenues, encouraging adoption by developers, maintaining service quality, and ensuring the economic viability of their infrastructure. Google AI Studio, Google’s development platform for its Gemini models, illustrates this tension with its current model combining a limited free offer and pay-per-use API access, but other players like OpenAI or Anthropic face similar trade-offs.

This paper proposes a radical alternative: what if a platform offered free access to its AI infrastructure and in return took a percentage of the revenues generated by developers' applications? This model, which we call "Revenue-Sharing as Infrastructure" (RSI), reverses the traditional upstream payment logic and creates an economic partnership between the platform and its complementors.

Our contribution is threefold. First, we provide a systematic review of the emerging literature on business models for GenAI platforms, identifying three successive generations of models. Second, we develop an original theoretical framework drawing on concepts of value co-creation, incentive mechanisms, and multi-layer market architecture to analyze the properties of the RSI model. Third, we propose a detailed technical architecture and discuss the conditions for its implementation.

The paper is organized as follows. Section 2 presents a literature review structured along a diachronic perspective. Section 3 develops the theoretical framework. Section 4 details the RSI proposal. Section 5 offers a comprehensive comparative analysis. Section 6 analyzes the societal impact of the model, especially for emerging economies. Section 7 discusses implications and challenges. Section 8 concludes and opens avenues for future research.

\section{State of the Art: The Evolution of Business Models for GenAI Platforms}

The literature on business models for generative AI platforms has been developing rapidly since 2023. We propose a diachronic reading identifying three successive generations of models, illustrated with simple analogies to facilitate understanding.

\subsection{First Generation: Pay-per-Use (like Gasoline)}

The first API offerings for generative AI, launched in late 2022 and early 2023, naturally adopted the business models of cloud computing. OpenAI, for instance, proposed a pay-per-token consumption model patterned after AWS or Azure. This model works like buying gasoline: you pay based on what you consume. Every API request, every token processed, is billed.

This model has the advantage of simplicity and predictability for the provider: the more developers use AI, the more the platform earns. However, it transfers the entire adoption risk to developers. If a developer creates an application that takes time to find its audience, they still have to pay API costs every month. This is a significant barrier for innovators without initial capital.

Research by Wang et al. shows that this first generation of models is characterized by a focus on the value proposition and key resource components of the Business Model Canvas, while revenue streams remain largely misaligned, indicating "uncertainty about commercialization paths." In short, they knew how to offer the service, but they did not yet know how to monetize it optimally.

\subsection{Second Generation: Diversification (Freemium and Subscriptions, like Spotify)}

From 2024 onward, a diversification of business models can be observed. Some platforms adopt vertical differentiation strategies: more powerful models at higher prices, faster models with specific rates. Others experiment with subscription plans for intensive developers. This generation can be compared to Spotify: there is a free version with limitations (ads, lower quality) and a premium version without limits. Similarly, AI platforms often offer a free tier with quotas (limited number of requests, less powerful models) and paid tiers with higher caps or advanced features.

The research by Yu et al. provides crucial insight into this period. Their game-theoretic analysis shows that "introducing a marketplace can sometimes hurt profitability" and that "providers' decisions about the quality of free services are influenced by the difficulty of creating customized services." In other words, finding the right balance between what is offered for free and what is charged is not straightforward. Too much free can attract users but also cannibalize potential revenues. These results suggest that optimizing business models for GenAI platforms is far from trivial and requires a careful consideration of incentives.

\subsection{Third Generation: The Emergence of Revenue Sharing (like YouTube)}

The most recent literature explores more complex market architectures, which can be likened to the YouTube model. On YouTube, content creators produce videos for free, the platform hosts and distributes them, and when videos generate advertising revenue, YouTube takes a cut (about 45\%) and gives the rest to the creator. Interests are aligned: the more successful a video, the more both the platform and the creator earn.

Ai et al. analyze "nonlinear procurement mechanisms" in two- or three-layer markets where platforms, human creators, and generative AI coexist. Their major contribution is to identify the growing role of "data intermediaries" (like Scale AI or creator organizations), which introduce a third layer into the traditional platform–creator structure. Their results show that "this three-layer market can lead to a lose-lose outcome, reducing both platform revenue and social welfare." Adding intermediaries can thus complicate matters and harm efficiency.

Meanwhile, Gao et al. study the impact of GenAI tool adoption on creative platforms. Their findings counter some intuitions: "although GenAI tools can facilitate content creation by enabling higher quality, creators' creativity levels and prices may decrease compared to the regime without these tools." They also show that the platform opts for "relatively weak regulation when AI intelligence is relatively low." In other words, AI can both help and harm depending on how it is used.

Our proposal, the RSI model, falls within this third generation but simplifies the architecture to two layers (platform and developers) and focuses on fair revenue sharing, without additional intermediaries. It is a kind of "YouTube for AI" where developers become application creators.

\subsection{Synthesis: Key Dimensions of GenAI Business Models}

The literature allows us to identify several structuring dimensions of business models for GenAI platforms:

\begin{enumerate}
    \item \textbf{Direction of financial flows}: upstream (developer to platform) vs. downstream (end user to platform)
    \item \textbf{Risk sharing}: risk borne by developer, platform, or shared
    \item \textbf{Incentive mechanisms}: how to align the interests of different stakeholders
    \item \textbf{Market architecture}: number of layers and relationships between them
    \item \textbf{Value co-creation}: mechanisms by which developers and platform jointly create value
\end{enumerate}

Table 1 summarizes the characteristics of the three identified generations.

\begin{table*}[h]
\centering
\caption{Evolution of Business Models for GenAI Platforms}
\label{tab:evolution}
\begin{tabular}{@{}p{3cm}p{3.5cm}p{3.5cm}p{3.5cm}@{}}
\toprule
\textbf{Dimension} & \textbf{First generation (2022-2023)} & \textbf{Second generation (2024-2025)} & \textbf{Third generation (emerging)} \\
\midrule
\textbf{Dominant model} & Pay-per-use (pay-per-token) & Freemium + differentiation & Multi-layer marketplaces \\
\textbf{Flow direction} & Upstream (developer → platform) & Mixed & Downstream potential \\
\textbf{Risk} & Entirely on developer & Partially mutualized & To be defined \\
\textbf{Incentives} & Linear (consumption) & Non-linear (quality, exclusivity) & Complex (revenue sharing) \\
\textbf{Complexity} & Low & Medium & High \\
\textbf{Key citations} & Wang et al. (2025) & Yu et al. (2025) & Ai et al. (2026), Keinan (2025) \\
\bottomrule
\end{tabular}
\end{table*}

\section{Theoretical Framework}

To analyze the properties of the RSI model we propose, we draw on three complementary theoretical frameworks.

\subsection{Value Co-Creation in Platform Ecosystems}

Heimburg, Schreieck, and Wiesche recently studied "complementor value co-creation in generative AI platform ecosystems." Their case study of OpenAI shows that value is not simply created by the platform alone, but emerges from the interaction between the platform and developers. They identify four concrete mechanisms by which developers contribute to this co-creation:

\begin{itemize}
    \item \textbf{System instructions}: developers fine-tune the behavior of the AI by giving it precise instructions, e.g., "answer in plain language" or "use a professional tone." This adapts the AI to their context.
    \item \textbf{Contextual data}: they provide domain-specific information \cite{22,23,24,25} (legal documents, medical data, product catalogs) so that the AI produces relevant and personalized answers.
    \item \textbf{User input curation}: they organize, filter, or reformulate user requests before sending them to the AI, improving the quality of responses.
    \item \textbf{Output revision}: they check, correct, or enrich the AI's responses before returning them to the user, ensuring service reliability.
\end{itemize}

These actions show that the final value of an application is not simply "extracted" from the platform, but is co-created by the joint work of the platform and the developer. Heimburg et al. also distinguish two logics of action among developers:
- \textbf{Reap logic}: the developer takes advantage of the existing capabilities of the AI without seeking to modify them deeply. They use the AI "as is" to meet standard needs.
- \textbf{Differentiation logic}: the developer customizes the AI to stand out and offer a unique service, investing in the co-creation mechanisms above.

This framework is particularly relevant for our proposal because it emphasizes that value is co-created. A business model that aligns the incentives of both parties (platform and developer) could therefore enhance these co-creation mechanisms. In the RSI model, the platform has every interest in encouraging developers to use these mechanisms, since its revenue depends on the success of the applications.

\subsection{Incentive Mechanisms and Revenue Sharing}

Keinan proposes a game-theoretic model analyzing strategic interactions between a platform and its content creators in the presence of generative AI. The originality of his approach lies in capturing "creators' dual strategic decisions: their investment in content quality and their (possible) consent to share their content with the platform's AI." To incentivize creators, the platform strategically allocates a portion of its AI-generated revenues to those who share their content.

This work is fundamental to our proposal because it mathematically demonstrates the possibility of equilibria where all creators voluntarily choose to share their content ("full-sharing equilibrium profiles"). Keinan establishes a "surprising link to the prisoner's dilemma" and shows how revenue allocation mechanisms affect creators' utility and platform revenues. Essentially, a well-designed revenue-sharing scheme can lead to a situation where everyone gains from cooperating.

\subsection{The Architecture of Multi-Layer Markets}

Ai et al. propose an analysis of "nonlinear procurement mechanisms" in markets where platforms, human creators, and generative AI coexist. Their major contribution is to identify the conditions under which introducing a third layer (data intermediaries) can degrade market efficiency. They show that intermediaries like Scale AI or creator organizations can, through pre-signed contracts, distort the incentives of individual creators and lead to a lose-lose situation for the platform and society.

This framework allows us to situate our proposal in a two-layer architecture (platform $\leftrightarrow$ developers) with a revenue-sharing mechanism that internalizes cross-layer externalities. By avoiding intermediaries, RSI aims to preserve efficiency and maximize the value created.

\subsection{Formal Modeling of RSI}
We propose a mathematical formalization of the RSI model to analyze its properties and determine the conditions for equilibrium. This modeling is inspired by Keinan's work \cite{keinan2025} on revenue-sharing mechanisms and two-sided platform theory.

\subsubsection{Actors and Parameters}
Consider a platform (e.g., Google, OpenAI) offering free AI infrastructure to a set of developers $D$. Each developer $i \in D$ develops an application generating revenue $R_i$ from end users. The platform takes a commission $\alpha \in [0,1]$ on this revenue, returning $(1-\alpha)R_i$ to the developer. The marginal cost borne by the platform for each API request is $c$ (technical cost). The total number of requests made by application $i$ is $q_i$, linked to revenue by a function $R_i = f_i(q_i)$. We assume this function is \textsc{increasing}: the more the application is used (i.e., the higher $q_i$), the higher the generated revenue. In other words, demand for the application translates into revenue growth as usage increases. Furthermore, we assume this function is \textsc{concave}, reflecting the idea of \textsc{diminishing returns}: beyond a certain level of usage, each additional request generates a smaller revenue gain. This can be explained by saturation effects (e.g., first users are willing to pay a high price, but to attract additional users one must lower prices or face increased competition) or increasing acquisition costs. Mathematically, concavity means that $f_i'(q_i)$ is positive but decreasing, a standard assumption in economic modeling for situations where increased usage yields decreasing marginal benefits.

\subsubsection{Developer's Decision}
The developer chooses the quality of their application (via an effort $e_i$) and the price $p_i$ charged to users. These choices affect both the number of users and token consumption. Revenue $R_i$ thus depends on effort and price: $R_i = R_i(e_i, p_i)$. For example, higher effort improves application quality, which can attract more users or allow a higher price, while price directly influences demand. The developer's net profit is:
\[
\pi_i = (1-\alpha)R_i(e_i, p_i) - \phi(e_i)
\]
where $\phi(e_i)$ is the cost of effort (strictly increasing and convex). The developer maximizes $\pi_i$ by choosing $e_i$ and $p_i$, leading to standard first-order conditions.

\subsubsection{Platform's Decision}
The platform chooses the commission rate $\alpha$ and, optionally, an investment in infrastructure quality (not modeled here). Its total profit is:
\[
\Pi = \sum_{i \in D} \left( \alpha R_i - c q_i \right)
\]
The platform anticipates developers' reactions to $\alpha$. The set of participating developers also depends on $\alpha$: a developer only participates if their profit $\pi_i$ exceeds a reservation value $\pi_0$ (outside option). Let $N(\alpha)$ be the number of participating developers.

\subsubsection{Game Equilibrium}
The equilibrium is a Stackelberg equilibrium: the platform sets $\alpha$ first, then developers choose their effort and price. Solving backwards yields the optimality conditions.

For a given $\alpha$, each developer $i$ maximizes $\pi_i$. The first-order condition with respect to effort gives:
\[
(1-\alpha) \frac{\partial R_i}{\partial e_i} = \phi'(e_i)
\]
This shows that effort is decreasing in $\alpha$: a higher commission reduces the incentive to invest.

The platform chooses $\alpha$ to maximize $\Pi(\alpha)$. Differentiating and accounting for the effects on $N(\alpha)$ and $R_i(\alpha)$ yields a trade-off condition between the number of developers and revenue per developer.

\subsubsection{Determining the Optimal Commission Rate}
In this subsection, we seek the optimal commission rate $\alpha^*$ that maximizes the platform's profit. To simplify, assume all developers are identical (same revenue and cost functions). The platform's profit becomes:
\[
\Pi(\alpha) = N(\alpha) \times \bigl[\alpha R(\alpha) - c q(\alpha)\bigr]
\]
where:
\begin{itemize}
    \item $N(\alpha)$ is the number of participating developers, decreasing in $\alpha$: a higher commission makes it less profitable for developers to join.
    \item $R(\alpha)$ is the revenue generated by each developer, which depends on $\alpha$ because the commission influences effort and thus application quality.
    \item $q(\alpha)$ is the number of requests per application, also a function of $\alpha$.
    \item $c$ is the unit cost borne by the platform for each API request.
\end{itemize}
The platform must trade off two opposing effects. On one hand, a higher commission yields more per developer (term $\alpha R$), but on the other hand, it reduces the number of developers $N$ (some drop out) and discourages effort, thereby decreasing $R$ and $q$. To find the optimal rate, set the derivative of $\Pi$ with respect to $\alpha$ to zero:
\[
\frac{d\Pi}{d\alpha} = N'(\alpha)(\alpha R - c q) + N(\alpha)\left( R + \alpha \frac{dR}{d\alpha} - c \frac{dq}{d\alpha} \right) = 0
\]
The first term, $N'(\alpha)(\alpha R - c q)$, is negative because $N'(\alpha)<0$ (fewer developers) and $\alpha R - c q >0$ (otherwise the platform would lose money on each developer). It represents the loss due to reduced participation. The second term, $N(\alpha)\left( R + \alpha \frac{dR}{d\alpha} - c \frac{dq}{d\alpha} \right)$, captures the effect on profit per developer: an increase in $\alpha$ directly raises the share taken ($R$), but it reduces $R$ and $q$ through developers' reactions.

\paragraph{Illustrative Example.}
To better understand the mechanism, consider a simple functional form. Assume a developer's revenue is proportional to effort and the cost of effort is quadratic: $R = e$, $\phi(e) = \frac{1}{2}e^2$. Also assume the number of requests is proportional to effort: $q = e$ (the better the application, the more interactions). Finally, simplify participation by setting $N(\alpha) = 1$ (consider a representative developer), isolating the effect on effort.

The developer maximizes $(1-\alpha)e - \frac{1}{2}e^2$. The first-order condition gives $1-\alpha - e = 0$, so $e = 1-\alpha$. Then $R = 1-\alpha$ and $q = 1-\alpha$. The platform's profit is:
\[
\Pi(\alpha) = \alpha(1-\alpha) - c(1-\alpha) = (1-\alpha)(\alpha - c)
\]
For $c < 1$, this profit is a concave parabola in $\alpha$. Setting the derivative to zero yields:
\[
\begin{aligned}
\frac{d\Pi}{d\alpha} &= (1-\alpha) - (\alpha - c) = 0 \\
&\quad\Rightarrow\quad 1 - 2\alpha + c = 0 \\
&\quad\Rightarrow\quad \alpha^* = \frac{1+c}{2}
\end{aligned}
\]

Thus, the optimal commission rate increases with the cost $c$ borne by the platform. For instance, if $c = 0.2$, then $\alpha^* = 0.6$; if $c = 0.4$, then $\alpha^* = 0.7$. The higher the infrastructure cost, the more the platform must charge to cover its costs, but this reduces developer effort. This simple result illustrates the fundamental trade-off in the model.

This example can be generalized by introducing demand elasticity (e.g., $R = A e^{\beta}$) and a participation function $N(\alpha)$ that accounts for competition among developers. It can then be shown that $\alpha^*$ is larger when the marginal cost $c$ is low and when demand is elastic (i.e., users are price sensitive), confirming the intuition above.

\subsubsection{Equilibrium Properties}
\begin{itemize}
    \item \textbf{Effect of commission on innovation}: Effort $e^*$ is decreasing in $\alpha$. A commission that is too high reduces application quality.
    \item \textbf{Developer participation}: The profitability threshold determines $N(\alpha)$. If the distribution of $\pi_0$ is known, $N(\alpha)$ can be computed.
    \item \textbf{Comparison with pay-per-token model}: In the traditional model, the developer pays $c q$ to the platform, a certain cost. Comparing expected profits shows that the RSI model is preferred by developers with low capital, but may be less advantageous for those with low effort costs.
\end{itemize}

\subsubsection{Possible Extensions}
The model can be extended to include:
\begin{itemize}
    \item \textbf{Heterogeneity of developers}: different types (skills, target markets).
    \item \textbf{Platform competition}: multiple choices for developers.
    \item \textbf{Network externalities}: the success of an application attracts users, increasing demand for infrastructure.
\end{itemize}

\section{Proposal: The Revenue-Sharing as Infrastructure (RSI) Model}

\subsection{Conceptual Foundations}

The RSI model we propose is based on a fundamental inversion of the economic logic of GenAI platforms. Instead of charging for access to the infrastructure upstream, the platform offers this access for free and is compensated through a percentage of the revenues generated by developers' applications.

This model draws inspiration from:
\begin{itemize}
    \item Application marketplaces (App Store, Google Play) that charge a commission on sales.
    \item Subscription models with revenue sharing (e.g., Substack).
    \item The incentive mechanisms identified by Keinan.
\end{itemize}

\subsection{Operational Principles}

The RSI model rests on three operational principles:

\textbf{Principle 1: Free access to AI infrastructure.} The platform no longer charges for API usage. Developers can create and deploy applications without any variable cost related to token consumption.

\textbf{Principle 2: Mandatory integration of the platform's payment system.} Any application using the free infrastructure must integrate the payment system provided by the platform (e.g., Google Pay, Stripe Connect, etc.) to monetize its services with end users.

\textbf{Principle 3: Revenue sharing.} The platform takes a percentage (e.g., 20–30\%) of each transaction made through the application. The remainder is remitted to the developer.

\textbf{Principle 4 (optional): Advertising monetization.} To diversify revenue sources and offer an alternative to applications that do not generate direct transactions, the platform can integrate an advertising network. Advertising revenue is shared according to the same principle: a percentage (e.g., 20–30\%) for the platform, the rest for the developer. This mechanism allows monetizing the audience even in the absence of user payments, while strengthening the economic viability of the model.

Figure 1 illustrates the architecture of this model.

\begin{figure*}[h]
\centering
\begin{tikzpicture}[
    node distance=2.2cm,
    plateforme/.style={
        rectangle, 
        draw, 
        fill=blue!30!white, 
        minimum width=3.5cm, 
        minimum height=1.3cm, 
        rounded corners=0.3cm,
        drop shadow={shadow xshift=2pt, shadow yshift=-2pt, opacity=0.3},
        font=\bfseries\sffamily
    },
    developpeur/.style={
        rectangle, 
        draw, 
        fill=green!30!white, 
        minimum width=3cm, 
        minimum height=1.3cm, 
        rounded corners=0.3cm,
        drop shadow,
        font=\bfseries\sffamily
    },
    user/.style={
        rectangle, 
        draw, 
        fill=orange!30!white, 
        minimum width=3cm, 
        minimum height=1.3cm, 
        rounded corners=0.3cm,
        drop shadow,
        font=\bfseries\sffamily
    },
    paiement/.style={
        rectangle, 
        draw, 
        fill=purple!30!white, 
        minimum width=3.5cm, 
        minimum height=1.3cm, 
        rounded corners=0.3cm,
        drop shadow,
        font=\bfseries\sffamily
    },
    tech/.style={
        ->, 
        thick, 
        color=blue!80!black, 
        >=stealth,
        line width=1.2pt
    },
    mon/.style={
        ->, 
        thick, 
        dashed, 
        color=red!80!black, 
        >=stealth,
        line width=1.2pt
    },
    percent/.style={
        font=\small\sffamily,
        fill=white,
        inner sep=2pt,
        rounded corners=2pt,
        draw=gray!50,
        drop shadow={shadow xshift=1pt, shadow yshift=-1pt, opacity=0.2}
    },
    flowlabel/.style={
        font=\small\sffamily\itshape,
        fill=white,
        inner sep=2pt,
        rounded corners=2pt,
        draw=gray!30
    }
]

\node[developpeur] (dev) {\textbf{Developer}};
\node[plateforme, right=of dev] (platform) {\textbf{AI Platform}};
\node[user, right=of platform] (enduser) {\textbf{End user}};
\node[paiement, below=2.5cm of platform] (pay) {\textbf{Platform Payment System}};

\draw[tech] (dev) -- node[flowlabel, above] {Creation} (platform);
\draw[tech] (platform) -- node[flowlabel, right, pos=0.3] {Free AI infrastructure} (pay);
\draw[tech] (pay) -- node[flowlabel, below, yshift=1.0cm, xshift=2.5cm] {Full service} (enduser);
\draw[tech] (enduser) -- node[flowlabel, above, yshift=-1.0cm, xshift=-0.5cm] {Requests} (pay);

\draw[mon] (enduser) .. controls +(0,1) and +(1.5,2) .. node[flowlabel, above, sloped, pos=0.4] {Payment for service} (pay);
\draw[mon] (pay) .. controls +(-1,0) and +(0,-1) .. node[flowlabel, left, pos=0.7] {Commission} (platform);
\draw[mon] (pay) .. controls +(-1.5,-0.5) and +(0,-1.2) .. node[flowlabel, below, sloped, pos=0.5] {Payout} (dev);

\node[percent, above right=1cm and 3cm of pay] {Commission: 20-30\%};
\node[percent, below left=0.2cm and 0.5cm of pay] {Payout: 70-80\%};

\node[rectangle, draw, rounded corners, fill=white, drop shadow,
      below right=0.5cm and -1cm of platform, xshift=3cm, yshift=-3cm,
      minimum width=2.5cm, minimum height=1cm] (legend) {
    \begin{tabular}{l}
        \textcolor{blue!80!black}{$\longrightarrow$} Technical flow \\
        \textcolor{red!80!black}{$-\!\!-\!\!\rightarrow$} Monetary flow
    \end{tabular}
};

\begin{pgfonlayer}{background}
    \node[rectangle, draw, dashed, rounded corners=0.5cm, 
          fill=gray!5, inner sep=0.5cm, 
          fit=(dev)(platform)(pay)(enduser)(legend)] 
          (ecosystem) {};
\end{pgfonlayer}
\node[above=0.1cm of ecosystem, font=\large\sffamily\bfseries] {Platform Ecosystem};

\node[font=\scriptsize\sffamily, text=gray, above=0cm of dev] {app creator};
\node[font=\scriptsize\sffamily, text=gray, above=0cm of enduser] {client};
\node[font=\scriptsize\sffamily, text=gray, above=0cm of platform] {Gemini, GPT, Claude...};

\end{tikzpicture}
\caption{Proposed RSI model: architecture of technical (blue) and financial (red) flows. Percentages illustrate revenue distribution.}
\label{fig:rsi_model}
\end{figure*}
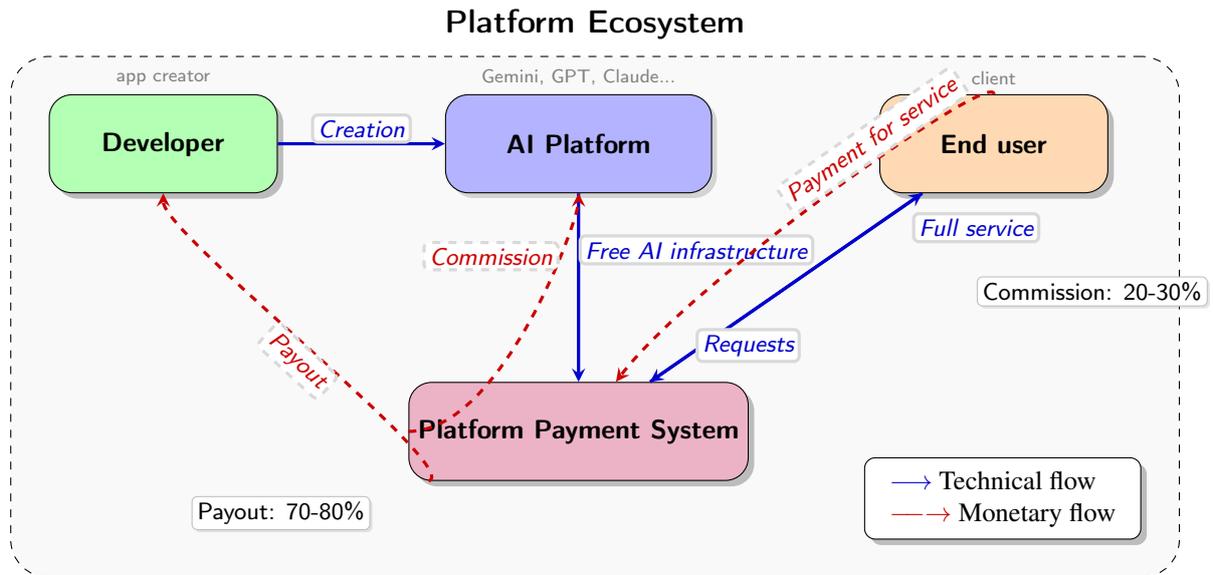

\subsection{Application Scenarios}

To illustrate the concrete functioning of the RSI model, consider three typical scenarios using the example of a platform like Google AI Studio, but applicable to other players.

\textbf{Scenario 1: Subscription-based legal advice application.} A developer creates an application using the platform's AI to analyze contracts and provide preliminary legal advice. The application is offered at \$20/month. With 1000 subscribers, monthly revenue is \$20,000. With a 25\% commission, the developer receives \$15,000 and the platform \$5,000.

\textbf{Scenario 2: Pay-per-image marketing application.} An image generation application charges \$0.50 per image. With 10,000 images generated per month, revenue is \$5,000, of which \$1,250 goes to the platform and \$3,750 to the developer.

\textbf{Scenario 3: Freemium application.} An application offers a limited free tier and a premium tier at \$10/month. Only premium users generate shared revenue, but all use the platform's infrastructure.

\section{Comparative Analysis}

\subsection{Methodology}

To rigorously evaluate the RSI model, we propose a multi-dimensional analysis comparing six business models identifiable in the literature:

\begin{enumerate}
    \item \textbf{Pay-per-token}: historical model
    \item \textbf{Freemium with limitations}: current model of many platforms (e.g., Google AI Studio)
    \item \textbf{Developer subscription}: monthly flat fee for unlimited access
    \item \textbf{Marketplace with commission}: platform connecting developers and users
    \item \textbf{Hybrid model}: combination of several approaches
    \item \textbf{RSI (our proposal)}: free infrastructure with revenue sharing
\end{enumerate}

\subsection{Dimensions of Analysis}

We analyze these models according to ten key dimensions derived from the literature and our theoretical framework. Table~\ref{tab:comparative} provides a synthetic comparative assessment.

\begin{table*}[h]
\centering
\caption{Multi-dimensional Comparative Analysis of Business Models}
\label{tab:comparative}
\resizebox{\textwidth}{!}{
\begin{tabular}{@{}lcccccc@{}}
\toprule
\textbf{Dimension} & \textbf{Pay-per-token} & \textbf{Freemium} & \textbf{Subscription} & \textbf{Marketplace} & \textbf{Hybrid} & \textbf{RSI (proposed)} \\
\midrule
\textbf{Entry barrier for developers} & High (variable costs) & Medium (quotas) & High (fixed cost) & Low & Medium & \textbf{None} \\
\textbf{Risk borne by developer} & High & Medium & High & Low & Medium & \textbf{Low} \\
\textbf{Risk borne by platform} & Low & Low & Low & High & Medium & \textbf{High} \\
\textbf{Alignment of interests} & Low & Low & Medium & High & Medium & \textbf{Very high} \\
\textbf{Incentive for innovation} & Constrained by costs & Medium & Medium & High & High & \textbf{Maximized} \\
\textbf{Technical complexity} & Low & Low & Low & High & Medium & \textbf{Medium} \\
\textbf{Contractual complexity} & Low & Low & Low & High & Medium & \textbf{High} \\
\textbf{Revenue potential for platform} & Function of usage & Function of usage & Fixed & Function of success & Mixed & \textbf{Function of success (unlimited upside)} \\
\textbf{Alignment with value co-creation} & Low & Low & Medium & High & Medium & \textbf{Maximal} \\
\textbf{Incentive equilibrium compatibility} & No & No & Partial & Yes & Partial & \textbf{Yes} \\
\bottomrule
\end{tabular}
}
\end{table*}

\subsection{Results of the Comparative Analysis}

We analyze each dimension of Table~\ref{tab:comparative}, justifying the evaluations and highlighting the relative positions of the different models, with a particular focus on the RSI model.

\paragraph{Entry barrier for developers.}
The \textbf{pay-per-token} model imposes immediate variable costs, a high barrier for unfunded developers. \textbf{Freemium} reduces this barrier via free quotas, but these are often insufficient for high-traffic applications. \textbf{Subscription} requires a fixed monthly payment, which can be prohibitive for small-scale projects. \textbf{Marketplace} lowers the barrier because the platform only charges upon success, but registration or maintenance fees may remain. The \textbf{hybrid} model combines several approaches and presents a medium barrier. \textbf{RSI} stands out with a zero barrier: the developer pays nothing upfront and bears no infrastructure costs, minimizing the entry threshold.

\paragraph{Risk borne by developer.}
In \textbf{pay-per-token} and \textbf{subscription} models, the developer bears a certain financial risk before generating any revenue. \textbf{Freemium} limits this risk via the free tier, but it remains when moving to the paid tier. \textbf{Marketplace} and \textbf{RSI} reverse the logic: risk is largely transferred to the platform, which only earns a commission if the application succeeds. The \textbf{hybrid} model presents medium risk as it may include both fixed and variable components.

\paragraph{Risk borne by platform.}
Symmetrically, traditional models impose low risk on the platform, as revenues are guaranteed by usage or subscriptions. \textbf{Marketplace} and \textbf{RSI} expose the platform to high risk: it must provide infrastructure without guaranteed revenue, betting on the success of applications. This risk is compensated by unlimited upside. Several mechanisms can mitigate this risk without altering the incentive logic. First, the platform can introduce an \textbf{activity threshold}: below a certain request volume, access remains free but costs are capped, preventing test-phase applications from incurring disproportionate charges. Second, a \textbf{degressive or progressive commission} can be applied (e.g., a higher rate for initial revenues, then reduced after a threshold), allowing more balanced risk sharing. The platform could also offer an optional \textbf{hybrid model}, letting developers choose between RSI and pay-per-use, limiting exposure to riskier projects. Moreover, integrating an \textbf{advertising module} (Principle 4) provides a complementary revenue source, independent of user transactions, diversifying income and offsetting infrastructure costs not covered by commissions alone. Finally, \textbf{risk pooling} across a large number of developers smooths costs: successes compensate failures, turning individual risk into a manageable portfolio risk. The \textbf{hybrid} model shares risk intermediately, and these mechanisms can be combined to enhance the platform's economic sustainability.

\paragraph{Alignment of interests.}
Alignment between platform and developers is crucial for a sustainable ecosystem. \textbf{Pay-per-token} and \textbf{freemium} create weak alignment because the platform has an incentive to maximize consumption (hence costs for the developer), while the developer seeks to minimize costs. \textbf{Subscription} offers medium alignment: once the subscription is paid, the platform has no further direct incentive to support the developer. \textbf{Marketplace} creates stronger alignment because the platform only gains if the developer sells. \textbf{RSI} achieves very high alignment: the platform and developer share the same goal of maximizing the application's revenue, motivating the platform to provide quality infrastructure and active support.

\paragraph{Incentive for innovation.}
Innovation is favored when developers can experiment without risk and are rewarded for success. \textbf{Pay-per-token} stifles innovation through variable costs. \textbf{Freemium} and \textbf{subscription} offer medium incentives. \textbf{Marketplace} encourages innovation via low barriers and success-based remuneration. \textbf{RSI} maximizes incentives by eliminating all initial costs and directly rewarding value creation, in line with Keinan's predictions \cite{keinan2025} on revenue-sharing mechanisms.

\paragraph{Technical complexity.}
Simple models like \textbf{pay-per-token} and \textbf{freemium} require only basic API integration. \textbf{Subscription} may require billing cycle management. \textbf{Marketplace} and \textbf{RSI} introduce higher technical complexity due to payment system integration and transaction tracking. \textbf{RSI} is rated medium complexity as the platform would provide ready-to-use SDKs and APIs for payment.

\paragraph{Contractual complexity.}
Contracts for \textbf{pay-per-token} and \textbf{freemium} are simple (standard ToS). \textbf{Subscription} may include commitment clauses. \textbf{Marketplace} requires partnership contracts detailing commissions. \textbf{RSI} demands high contractual complexity to define revenue sharing, audits, dispute resolution, and regulatory compliance.

\paragraph{Revenue potential for platform.}
\textbf{Pay-per-token} and \textbf{freemium} yield usage-based revenues, with potential ceiling but limited by developers' payment capacity. \textbf{Subscription} generates fixed, predictable but capped revenues. \textbf{Marketplace} and \textbf{RSI} allow unlimited upside, as revenues follow application success (multiplier effect). \textbf{RSI} combines this upside with a broader developer base, potentially generating higher long-term revenues.

\paragraph{Alignment with value co-creation.}
Heimburg et al.'s work \cite{heimburg2025} shows that value is co-created between platform and developers through instructions, contextual data, etc. Models where the platform is merely a service provider (\textbf{pay-per-token}, \textbf{freemium}) do not foster this co-creation. \textbf{Subscription} and \textbf{marketplace} encourage it partially. \textbf{RSI} is maximally suited because the platform has every interest in facilitating co-creation mechanisms to increase shared revenues.

\paragraph{Incentive equilibrium compatibility.}
Keinan \cite{keinan2025} showed that revenue-sharing mechanisms can lead to equilibria where all creators choose to share their content. \textbf{Pay-per-token} and \textbf{freemium} do not permit such equilibria. \textbf{Subscription} can partially contribute if the plan includes sharing. \textbf{Marketplace} is compatible with incentive equilibria. \textbf{RSI}, by its very principle of revenue sharing, fully satisfies this condition.

In summary, the RSI model stands out with a unique combination of low barriers, shared risk, high interest alignment, strong innovation incentives, and unlimited revenue potential. These properties make it a promising candidate to stimulate a dynamic and innovative ecosystem around generative AI platforms.

\section{Societal Impact of the RSI Model: Opportunities for Emerging Economies}

Beyond the direct economic implications for platforms and developers, the RSI model could play a catalytic role in the development of low-income countries and those facing high unemployment. By removing financial entry barriers, it opens the way for broader participation of innovators from regions hitherto marginalized in the digital economy. This section analyzes this societal impact across several dimensions.

\subsection{Digital Divide and Mobile Connectivity}

Analysis of the potential impact of the RSI model must account for the contrasting realities of digital infrastructures around the world. According to the International Labour Organization (ILO), workers in low-income countries face a paradox: "those who hold jobs vulnerable to automation generally have sufficient internet connectivity to suffer displacement effects, while those who could benefit from productivity gains from generative AI face substantial digital infrastructure gaps that could prevent them from realizing these productivity gains" \cite{gmyrek2026}.

Despite these disparities, encouraging data emerge. The 2025 Global Findex report from the World Bank reveals that 84\% of adults in low- and middle-income countries own a mobile phone, three-quarters of which are smartphones \cite{morgandi2026}. In these countries, 90\% of internet users access it via mobile, making the smartphone the primary gateway to income-generating digital services. This mobile penetration provides a basic infrastructure on which the RSI model could rely to democratize access to AI application development capabilities.

\subsection{Unlocking the "Latent Jobs Dividend"}

The World Economic Forum introduces the concept of "Latent Jobs Dividend" to describe the untapped economic potential in the Global South \cite{wef2025}. According to this analysis, "the problem is not the absence of work. It is that millions of jobs remain latent – blocked by distribution gaps, degree bottlenecks, and legacy systems." These jobs are not hypothetical: they correspond to essential services that populations are willing to pay for today.

The RSI model, by allowing developers without initial capital to create applications addressing these needs, could help unlock this potential. The cost of inaction is quantified in the same report:

\begin{itemize}
    \item \textbf{Health}: According to the World Health Organization, over 50\% of people in low-income countries lack access to essential health services \cite{wef2025}.
    \item \textbf{Agriculture}: Sub-Saharan Africa loses more than \$4 billion annually due to preventable crop and livestock diseases.
    \item \textbf{Legal services}: Over 90\% of African small businesses operate in the informal sector, often without proper contracts or access to affordable legal services.
\end{itemize}

A mother in Zambia may wait three days and lose \$30 in wages to reach a health center, while a mobile AI-powered tool could offer a diagnosis in minutes for less than a dollar. The RSI model would create the economic incentive for local developers to design such solutions.

\subsection{Sectoral Transformation and Job Creation}

\subsubsection{Agriculture}
Agriculture, which employs a significant share of the active population in developing countries, could greatly benefit from AI-based applications. Platforms like Hello Tractor already pilot tools for early detection of crop diseases, improving yields and food security \cite{wef2025}. The RSI model would allow agricultural entrepreneurs to develop advisory applications for farmers without bearing initial infrastructure costs, sharing the revenues generated by these services.

\subsubsection{Health}
Ethiopia has trained over 40,000 community health workers. With AI tools, these agents could "do more, faster, and with greater accuracy" \cite{wef2025}. Concrete applications already exist: a mother in Uganda can use her phone to record her newborn's first cries, and an app like Ubenwa instantly analyzes these cries to detect signs of birth asphyxia. The RSI model would encourage the development of such applications by aligning developers' and platforms' interests with measurable impact.

\subsubsection{Business Services and Formalization}
Lack of access to affordable legal and advisory services keeps many businesses in the informal sector. A young entrepreneur in Rwanda could use a legal AI accessible via SMS to analyze a lease and detect abusive clauses. These services, which would today require a law degree or a costly lawyer, become accessible through AI.

\subsubsection{Concrete Example: Jobop in Morocco}
The Moroccan startup Jobop illustrates the potential of digital platforms to structure the labor market in Africa. Founded in 2021, Jobop is the first 100\% digital temporary employment agency in Africa. It uses an advanced algorithm to match business needs with worker skills in real time \cite{afdb2025}. The temporary work market in Africa represents up to 65\% of the continent's employment, a market valued at \$100 billion. With annual growth over 30\%, Jobop has already raised \$1 million from a fund backed by the African Development Bank. This model demonstrates that digital platforms can structure entire sectors and create new economic opportunities.

\subsection{Effects by Continent and Region}

\subsubsection{Africa}
Africa has over 1.2 billion mobile connections, and platforms like M-Pesa and MTN MoMo prove that populations are ready to pay for accessible, immediate, and affordable services \cite{wef2025}. The African Development Bank emphasizes that "human capital is a key lever for African development" and that technology can be "a powerful accelerator for structuring employment." The RSI model could accelerate this dynamic by enabling a new generation of African developers to create applications addressing local needs, with no financial barrier to entry.

\subsubsection{South and Southeast Asia}
According to the World Bank, more than 10\% of adults in East Asia and the Pacific (excluding China) already earn money online \cite{worldbank2025}. Digital platforms like Shopee are transforming commerce, and 26\% of adults in low- and middle-income countries use the internet to learn, which can help them improve their services or enter new markets.

\subsubsection{Latin America}
In Mexico, Uber drivers reportedly earned about three to four times the official minimum wage (\$162 in 2018), net of vehicle rental costs \cite{unctad2025}. These earnings, while not guaranteeing social protection, demonstrate the potential of platforms to generate income superior to local alternatives.

\subsubsection{Europe and North America}
In mature economies, the RSI model could catalyze a wave of disruptive innovation by eliminating initial infrastructure costs. In Europe, where policies supporting digital entrepreneurship are numerous (notably through the Horizon Europe program), removing financial barriers would allow thousands of startups to quickly test their ideas without resorting to early fundraising. According to the European Commission, over 60\% of startups fail due to cash flow difficulties in the first two years; RSI would reduce this risk. In North America, where competition among AI platforms is intense, the model would offer a differentiating advantage for independent developers and small teams, allowing them to compete with better-capitalized players. Moreover, integrating an advertising module would meet the needs of applications with large audiences but low direct monetization, a common situation in media and online community sectors. Finally, formalizing revenues through the platform's payment system would facilitate tax and social compliance, aligning with regulatory requirements in these regions.

\subsection{Challenges and Success Conditions}

The positive impact of the RSI model will depend on implementing appropriate safeguards. The ILO emphasizes that AI adoption "varies across countries and sectors, depending not only on technological readiness but also on the enabling policy environment, institutional capacity, and social dialogue mechanisms" \cite{gmyrek2026}. Moreover, unlike informal businesses, "platforms are intrinsically traceable. They operate through a digital infrastructure, use formal payment systems, and generate rich data on transactions, hours worked, and worker performance" \cite{morgandi2026}. The RSI model, by integrating platform payment systems, would help formalize transactions and create traceability beneficial to both workers and authorities.

\subsection{Key Figures and Projections}

Table~\ref{tab:impact} summarizes key indicators illustrating the potential of the RSI model in emerging economies.

\begin{table*}[h]
\centering
\small
\caption{Key Indicators for the Societal Impact of the RSI Model}
\label{tab:impact}
\begin{tabular}{@{}p{9cm}cl@{}}
\toprule
\textbf{Indicator} & \textbf{Value} & \textbf{Source} \\
\midrule
Adults owning a mobile (low/middle-income countries) & 84\% & World Bank \cite{morgandi2026} \\
Mobile connections in Africa & 1.2 billion & WEF \cite{wef2025} \\
Share of temporary employment in Africa & up to 65\% & African Development Bank \cite{afdb2025} \\
Temporary work market in Africa & \$100 billion & African Development Bank \cite{afdb2025} \\
Adults earning money online (East Asia/Pacific) & >10\% & World Bank \cite{worldbank2025} \\
Adults using the internet to learn & 26\% & World Bank \cite{worldbank2025} \\
Inaccessible health services (low-income countries) & >50\% & WHO via WEF \cite{wef2025} \\
Annual agricultural losses (Sub-Saharan Africa) & \$4 billion & WEF \cite{wef2025} \\
\bottomrule
\end{tabular}
\end{table*}

\section{Discussion}

\subsection{Implications for Platforms}

The RSI model offers several strategic advantages for a platform like Google AI Studio, but also for other players (OpenAI, Anthropic, etc.):

\textbf{Ecosystem expansion.} By removing variable costs, the platform could attract a new category of developers: students, individual entrepreneurs, small businesses, innovators from emerging countries.

\textbf{Potentially higher revenues.} In a market where successful applications generate considerable revenue (e.g., SaaS apps), a 20-30\% commission could exceed the revenue of a pay-per-token model, especially if the volume of applications increases significantly.

\textbf{Strategic positioning.} The platform would position itself not merely as an API provider but as a genuine economic partner to developers, strengthening loyalty to its ecosystem.

\textbf{Competitive advantage.} This model could be hard for competitors with lower margins to copy, creating a strategic entry barrier.

\subsection{Implications for Developers}

\textbf{Advantages:}
- Zero initial financial risk
- Ability to test multiple ideas at no cost
- Access to cutting-edge infrastructure without investment
- Alignment with value co-creation logic

\textbf{Potential drawbacks:}
- Increased dependency on the chosen platform
- Revenue share taken (compared to API costs avoided)
- Contractual and technical complexity

\subsection{Challenges and Risks}

\textbf{Technical challenges:}
- Implementing a reliable transaction tracking system
- Fraud detection (under-reporting of revenues)
- Seamless integration of the payment system into applications

\textbf{Economic challenges:}
- Determining the optimal commission rate (too high discourages, too low insufficient)
- Managing applications that generate revenue outside the platform's payment system
- Balancing with existing offers (pay-per-use) during the transition

\textbf{Legal challenges:}
- Complex partnership contracts
- Compliance with marketplace regulations (e.g., DMA in Europe)
- Managing disputes over revenue calculations

\subsection{Limitations and Future Research}

Our study has several limitations that open avenues for further research:

\textbf{Empirical validation.} Our proposal remains theoretical. Empirical validation could involve interviews with developers, econometric modeling, or technical prototyping.

\textbf{Optimization of the commission rate.} Game-theoretic research could determine the optimal commission rate based on the elasticity of application supply and user demand.

\textbf{Cross-platform comparative analysis.} Applying the RSI model to different platforms (OpenAI, Anthropic, Mistral) could be studied, considering their specific technical and economic characteristics.

\textbf{Regulation.} The potential impact of regulations (DMA, DSA) on this model deserves in-depth analysis.

\textbf{Societal impact.} Empirical studies in developing countries would be needed to quantify the effects of the RSI model on employment, digital inclusion, and economic development.

\section{Conclusion}

This paper has proposed an original business model for generative AI platforms: "Revenue-Sharing as Infrastructure." By reversing the traditional upstream payment logic, this model lowers entry barriers for developers, aligns stakeholder interests, and enhances the value co-creation mechanisms identified in recent literature.

Our comparative analysis shows that this model, although more complex to implement, has unique properties that could make it a powerful growth lever for platforms like Google AI Studio, OpenAI, or other players in the sector. It fits within the third generation of GenAI business models, characterized by multi-layer market architectures and risk-sharing mechanisms.

The societal impact analysis further reveals the transformative potential of the RSI model for emerging economies. By leveraging high mobile penetration and unlocking the "latent jobs dividend," it could help reduce inequalities in access to essential services, formalize economic activities, and create employment opportunities in key sectors such as health, agriculture, and business services.

Future research is needed to empirically validate these propositions, optimize the model parameters, and analyze its competitive, regulatory, and societal implications. As generative AI profoundly transforms the digital economy, the question of business models for the platforms that host it remains a research field as crucial as it is nascent.

\section*{Acknowledgments}
The author thanks the researchers whose recent work has nourished this reflection, particularly the teams at ICIS 2025, HICSS 2025, and the contributors to the journals Technovation and Journal of Management Information Systems. He also wishes to express his gratitude to the Google AI Studio and Gemini team for providing an AI-based development platform, whose exploration directly inspired this proposal. He hopes that this model may one day be applied within this ecosystem, contributing to a more inclusive and innovative development.

\end{document}